\documentclass[a4paper,fleqn]{cas-dc}
\usepackage{threeparttable}
\usepackage{tablefootnote}
\usepackage{lineno,hyperref}
\usepackage{color}
\usepackage{multirow}
\usepackage{url}
\usepackage{comment}
\usepackage[authoryear,longnamesfirst]{natbib}
\usepackage[normalem]{ulem}

\begin{document}
\let\WriteBookmarks\relax
\def\floatpagepagefraction{1}
\def\textpagefraction{.001}

\shorttitle{
Two-component jet model for VHE GRBs
}   

\shortauthors{Sato et al. 2024}


\title[mode = title]{
Two-component jet model for the afterglow emission of GRB~201216C and GRB~221009A and implications for jet structure of very-high-energy gamma-ray bursts
}

\author[1]{Yuri~Sato}[orcid=0000-0003-2477-9146]
\ead{yuri@astr.tohoku.ac.jp}
\cormark[1]


\author[2,3]{Kohta~Murase}

\author[4]{Yutaka~Ohira}

\author[5]{Susumu~Inoue}

\author[6,7]{Ryo~Yamazaki}

\address[1]{Astronomical Institute, Graduate School of Science, Tohoku University, Sendai 980-8578, Japan}
\address[2]{Department of Physics; Department of Astronomy \& Astrophysics; Center for Multimessenger Astrophysics, Institute for Gravitation and the Cosmos, The Pennsylvania State University, University Park, PA 16802, USA}
\address[3]{Center for Gravitational Physics, Yukawa Institute for Theoretical Physics, Kyoto, Kyoto 606-8502, Japan}
\address[4]{Department of Earth and Planetary Science, The University of Tokyo, 7-3-1 Hongo, Bunkyo-ku, Tokyo 113-0033, Japan}
\address[5]{International Center for Hadron Astrophysics, Chiba University, 1-33 Yayoi-cho, Inage-ku, Chiba City, Chiba 263-8522}
\address[6]{Department of Physical Sciences, Aoyama Gakuin University, 5-10-1 Fuchinobe, Sagamihara, Kanagawa 252-5258, Japan}
\address[7]{Institute of Laser Engineering, Osaka University, 2-6, Yamadaoka, Suita, Osaka 565-0871, Japan}

\begin{abstract}
In recent years, afterglow emission in the very-high-energy (VHE) band above 100 GeV has been clearly detected for at least five gamma-ray bursts (GRBs~180720B, 190114C, 190829A, 201216C and 221009A).
For some of these VHE GRBs, we previously proposed a two-component jet model, consisting of two uniform jets with narrow and wide opening angles to explain their multiwavelength afterglows including VHE gamma rays.
In this paper, we show that the VHE spectra and light curves of GRBs~201216C and 221009A can also be reasonably explained by our two-component jet model, based on two top-hat jets propagating into a constant-density circumburst medium.
We find that for the five VHE GRBs, the collimation-corrected kinetic energies of the narrow and wide jets have typical values of $5\times10^{49}$~erg and $5\times10^{50}$~erg, respectively.
We discuss the similarities and differences among the VHE GRBs, and the implications for the structure of their jets. 
In agreement with previous studies, the narrow jet of GRB~221009A has an atypically small opening angle, so that its intrinsic, collimation-corrected energy remains within a plausible range despite the unusually large isotropic-equivalent energy.
\end{abstract}

 
\ExplSyntaxOn
\keys_set:nn { stm / mktitle } { nologo }
\ExplSyntaxOff

\maketitle


\section{Introduction}
\label{sec:intro}

Gamma-ray bursts (GRBs) are the most electromagnetically luminous transient phenomena in the Universe.
Some kinds of bursts (so called, long GRBs in ordinary nomenclature) are triggered by certain types of massive star core collapse events that generate relativistic jets \citep[e.g.][]{Piran2004,Zhang2019}.
Their prompt emission of primarily MeV-band photons, characterized by rapid, irregular variability and typical durations of seconds to minutes, is thought to arise from within the inner parts of the jets.
This is accompanied by afterglow emission that spans the radio to gamma-ray bands and decays more gradually over timescales of hours to days, understood to be radiated by nonthermal electrons accelerated in blastwaves that result from the interaction of the jets with the ambient medium.
Crucial new information on GRBs has been brought forth by the recent discovery of gamma rays in the very-high-energy (VHE) band above 100 GeV, achieved so far for five~GRBs,~GRB~180720B \citep{HESS2019}, GRB~190114C \citep{MAGIC2019b,MAGIC2019a}, GRB 190829A \citep{HESS2021}, GRB~201216C \citep{MAGIC2024} and GRB~221009A \citep{LHAASO2023a}.
The VHE emission of all bursts are observed to decay as power-laws in time, and thus most likely associated with the afterglow.
For the VHE radiation mechanism, synchrotron self-Compton (SSC), external inverse-Compton and hadronic processes have been proposed \citep[e.g.][]{Nava2021,Gill2022,Miceli2022}.
However, the details of the physics of VHE emission remain uncertain.
In the near future, the number of detectable VHE GRBs is expected to increase rapidly with current and future detectors like the Cherenkov Telescope Array (CTA) \citep{Kakuwa2012,Gilmore2013,Inoue2013,Sato2023a}.
VHE gamma-ray emission is expected to provide new insights to the physics of GRBs, including radiative processes, particle acceleration, dynamics of relativistic jets and the central engine.

The observed multiwavelength afterglows of the VHE GRBs have complicated light curves and interesting features in various bands.
The X-ray and optical emission of GRBs~180720B, 190114C, 201216C and 221009A are significantly brighter than that of GRB~190829A \citep{Ror2023}.
In the radio band, the luminosities of GRBs~190829A and 221009A are comparable, and are much lower than those of GRBs~180720B, 190114C and 201216C \citep{Rhodes2020,Laskar2023,Ror2023}.
Remarkably, the VHE gamma-ray light curve of GRB~221009A reveals a clear break where the light curve steepens at a few 100 seconds after the burst \citep{LHAASO2023a}.
This is most plausibly identified as a jet break \citep{Sari1999}, but the early break time implies a very narrow jet, which is inadequate to explain the emission observed at later times in other wavelengths \citep{LHAASO2023a}.
Furthermore, for GRB~221009A, the highest photon energy reached $\approx 10$~TeV \citep{LHAASO2023b}.
Such complicated multiwavelength afterglows are difficult to explain in the standard afterglow model assuming a single jet with a top-hat-shaped angular profile of the kinetic energy and bulk Lorentz factor.

We have shown 
in previous studies \citep{Sato2021,Sato2023a,Sato2023b}
that multiwavelength afterglows from some VHE GRBs can be explained by a two-component jet model consisting of two top-hat jets with different opening angles (i.e., ``narrow'' and ``wide'' jets) 
\citep{Ramirez-Ruiz2002}.
Such models were motivated for explaining the multiwavelength 
afterglow emission of some earlier GRBs \citep[e.g.,][]{Berger2003,Huang2004,Peng2005,Wu2005,Racusin2008,Rhodes2022}.
In constrast to top-hat jets,
hydrodynamic simulations have shown that
GRB jets could have angular structure with kinetic energy and bulk Lorentz factor gradually decreasing with angle from the jet axis
\citep[e.g.][]{Zhang2009,Gottlieb2021,Urrutia2023}.
The two-component jets proposed here may be considered the simplest approximation of such structured jets.

In this paper, we compare our predicted VHE gamma-ray emission from the narrow and wide jets determined by our previous studies with the updated observational results in the VHE gamma-ray bands of GRBs~201216C and 221009A.
The detailed VHE gamma-ray data from GRBs~201216C and 221009A had not yet been made public when our previous papers with the two-component jet model \citep{Sato2023a,Sato2023b} were published. In particular, in \citet{Sato2023b}, the model was compared only with an estimate of the number of VHE gamma-ray photons detected by the Large High Altitude Air Shower Observatory (LHAASO) as reported in the GCN Circular \citep{LHAASO2022}, instead of the full light curve and spectra that did not become available until later.
The paper is organized as follows.
In Section~\ref{sec:model}, we describe our model.
In Section~\ref{sec:result}, our model is compared with the observed afterglows of GRBs~201216C and 221009A. 
Section~4 is devoted to a discussion.
In section~\ref{subsec:compara}, 
our model for GRB~221009A
is compared with the other structured jet models proposed so far.
In Sections~\ref{subsec:common} and \ref{subsec:difference}, 
we summarize the similarities and differences among VHE GRBs, respectively.
In Section~\ref{subsec:implication}, we discuss the implications of our results. 
The conclusions are presented in Section~5.


\section{Model description of afterglow emission from relativistic jets}
\label{sec:model}

We utilize the basic formalism for the standard afterglow model, as done in our
previous works
\citep{Sato2021,Sato2023a,Sato2023b}. 
The afterglow emission from the two-component jets is given by a superposition of the radiation from two uniform (top-hat) jets. 
We assume that the central engine launches two jets simultaneously.
The external forward-shock emission from each jet component is described as follows \citep[see][for details]{Huang2000}.
Each relativistic jet component has initial isotropic-equivalent kinetic energy $E_{\rm{iso,K}}$, 
initial bulk Lorentz factor $\Gamma_0$ and initial opening half-angle $\theta_0$, and
propagates into a uniform circumburst medium (CBM) with number density $n_0$.
Note that in \citet{Huang2000}, the jet spreads sideways as it expands.
A thin shell is formed by the interaction between the jet and the surrounding matter, corresponding to the emission region. 
To calculate the synchrotron and SSC emission, we assume that the electron energy distribution at injection is a power-law with spectral index $p$, and that the microphysical parameters are time-independent, such as $\epsilon_e$, $\epsilon_B$ and $f_e$, 
which are the energy fractions of the internal energy going into non-thermal electrons, magnetic fields and the number fraction of accelerated electrons, respectively.
The flux density $F_{\nu}$ is obtained by integrating the emissivity over the equal arrival time surface \citep[e.g.][]{Granot1999}.
The SSC flux takes into account the Klein-Nishina effect self-consistently 
\citep[e.g.][]{Nakar2009,Murase2010,Murase2011,Wang2010,Jacovich2021,BTZhang2021,Sato2023a}.
We assume an on-axis viewing angle ($\theta_v=0$), which is justified by the high observed luminosity of the prompt emission of GRBs~201216C and 221009A. 
In this paper, we do not consider 
in detail
the external reverse-shock emission for simplicity.


\section{Results of afterglow modeling}
\label{sec:result}

We show the results of our modeling of the multiwavelength afterglow emission for GRB~201216C and GRB~221009A
in~\S\ref{subsec:GRB201216C} and \S\ref{subsec:GRB221009A}, respectively.

\subsection{GRB~201216C}
\label{subsec:GRB201216C}

\begin{figure*}
\centering
\begin{minipage}{0.31\linewidth}
\centering
\includegraphics[height=1.0\textwidth]{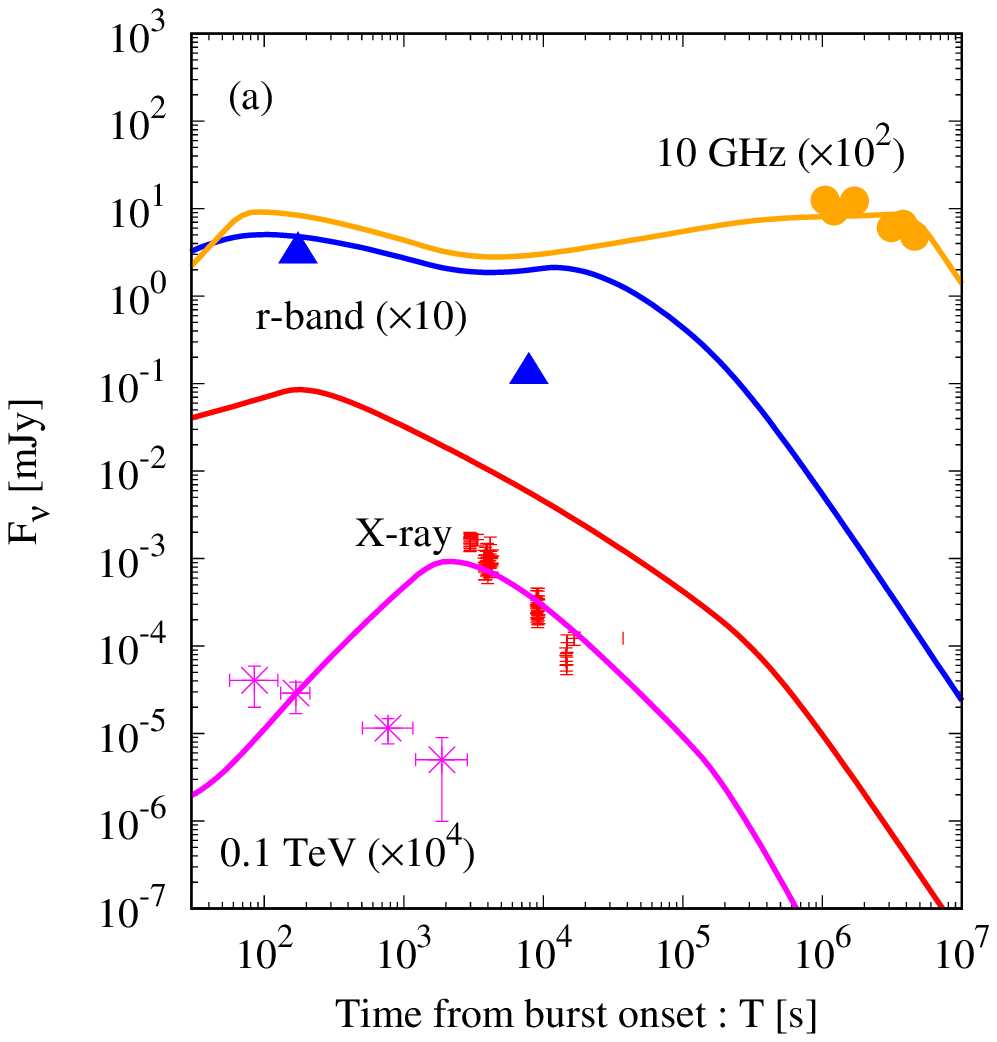}
\end{minipage}
\begin{minipage}{0.31\linewidth}
\centering
\includegraphics[height=1.0\textwidth]{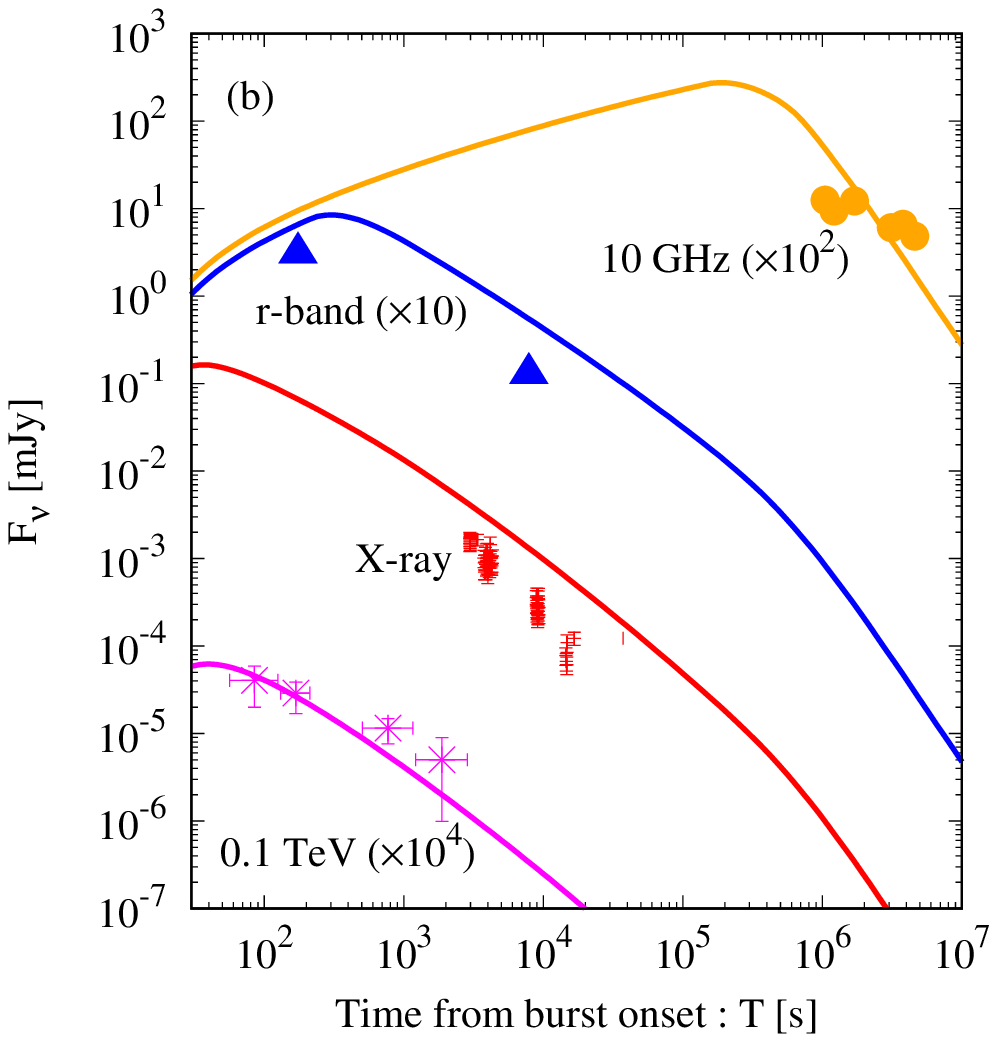}
\end{minipage}
\caption{
Multiwavelength afterglow light curves
of GRB~201216C for the single-component jet model
(VHE (0.1~TeV: magenta), X-ray (5~keV: red), optical (r-band: blue) and radio bands (10~GHz: orange)), 
compared with
observed data (VHE (0.1~TeV: magenta points), X-ray (5~keV: red points), optical (r-band: blue upward triangles) and radio bands (10~GHz: orange circles)).
Panels (a) and (b) show the models with the failed parameter sets~I and II, respectively (see text for details).
}
\label{fig:GRB201216Cf}
\end{figure*}

\begin{figure*}
\centering
\begin{minipage}{0.3\linewidth}
\centering
\includegraphics[height=1.0\textwidth]{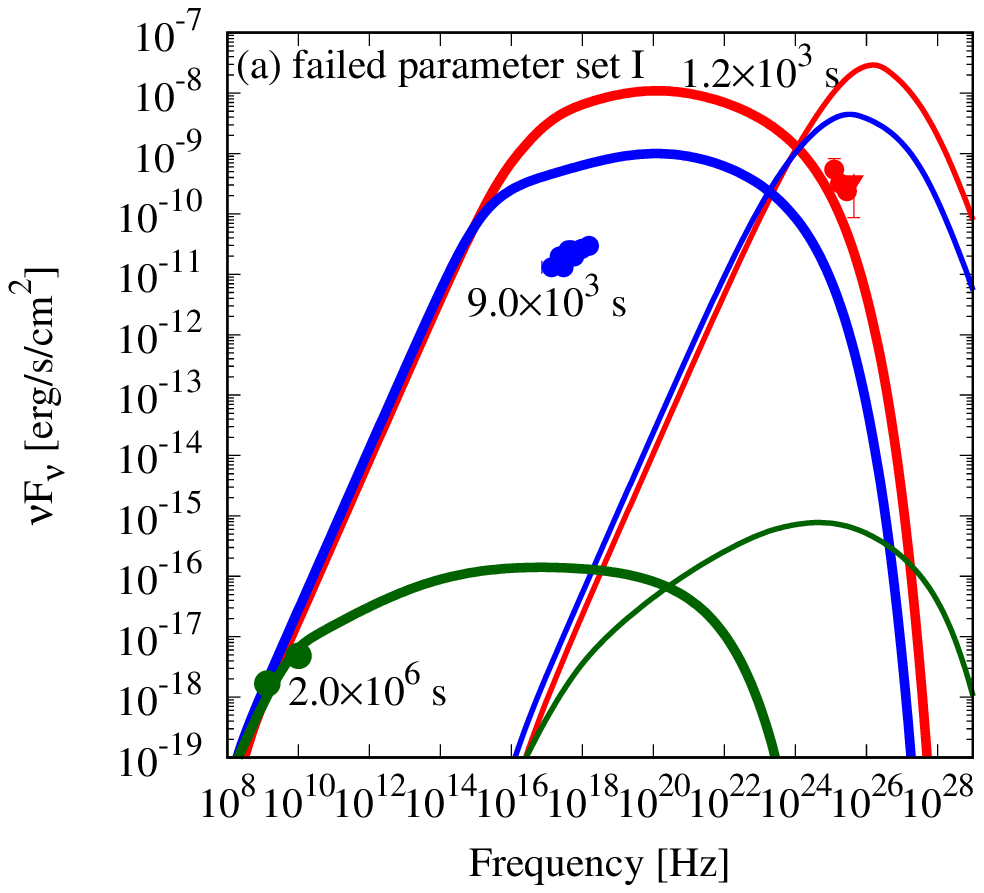}
\end{minipage}
\hspace{0.03\columnwidth}
\begin{minipage}{0.3\linewidth}
\centering
\includegraphics[height=1.0\textwidth]{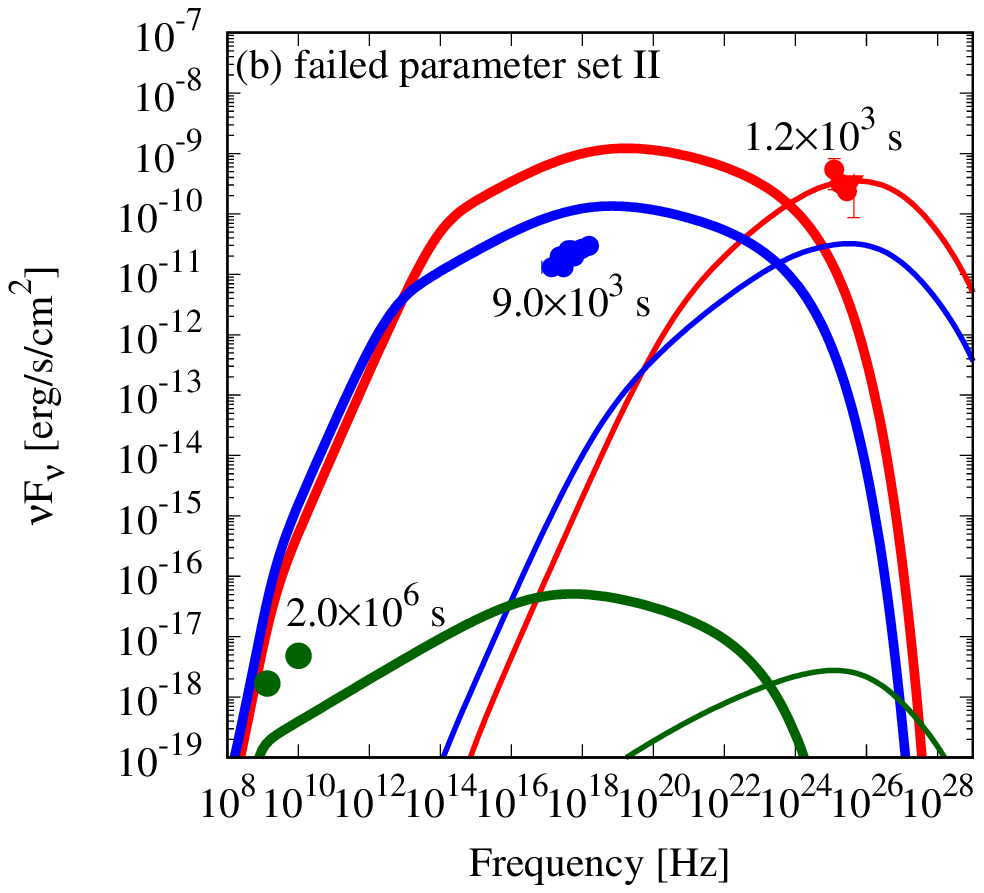}
\end{minipage}
\hspace{0.03\columnwidth}
\begin{minipage}{0.3\linewidth}
\centering
\includegraphics[height=1.0\textwidth]{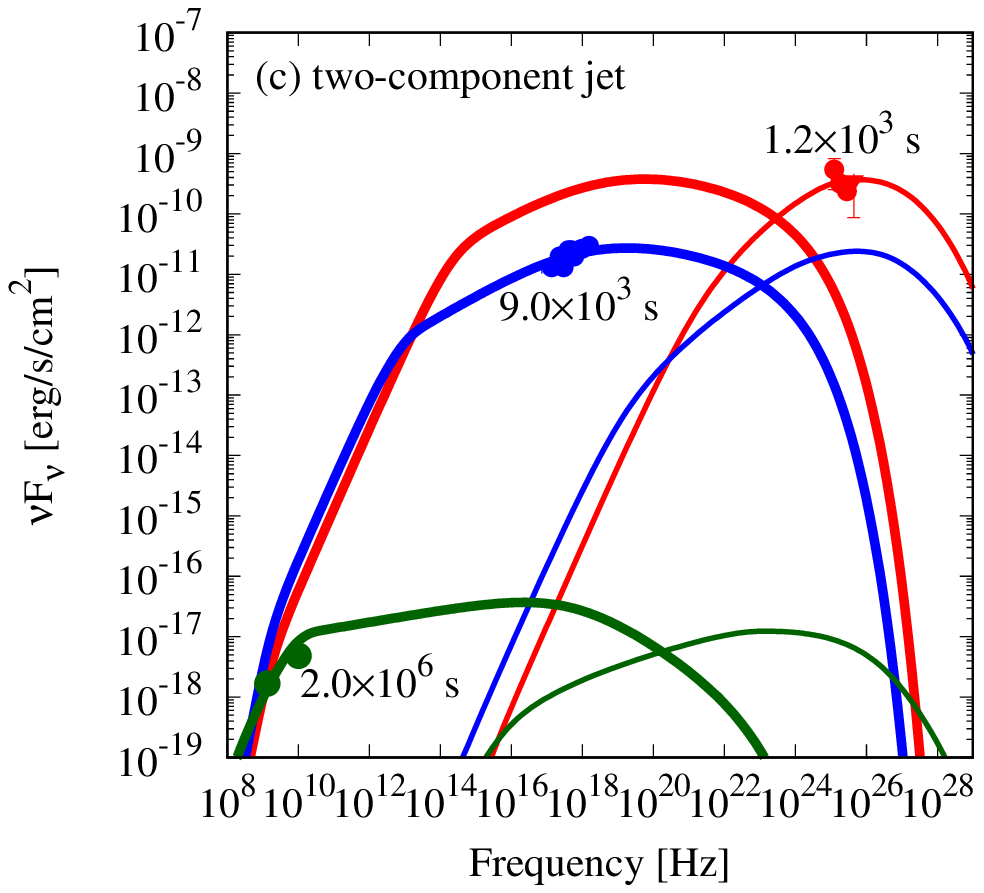}
\end{minipage}
\caption{
Model spectra of GRB~201216C for different
observer times ($T=1.2\times10^3$~s in red, $9.0\times10^3$~s in blue and $2.0\times10^6$~s in green), 
compared with observed data.
Thick and thin solid curves show the synchrotron and SSC  spectra, respectively.
Panel (a): result for the failed parameter set~I.
Panel (b): result for the failed parameter set~II.
Panel (c): result for the two-component jet model.
Note that in panel (c), the solid lines show the sum of the narrow and wide jets.
The narrow jet emission dominates at $T\sim1.2\times10^3$~s and $T\sim9.0\times10^3$~s (red and blue lines in panel (c)), 
while the wide jet emission becomes non-negligible at $T\sim2.0\times10^6$~s (green lines in panel (c)).
}
\label{fig:GRB201216C_SED}
\end{figure*}

\begin{figure}
\centering 
\includegraphics[width=0.31\textwidth]{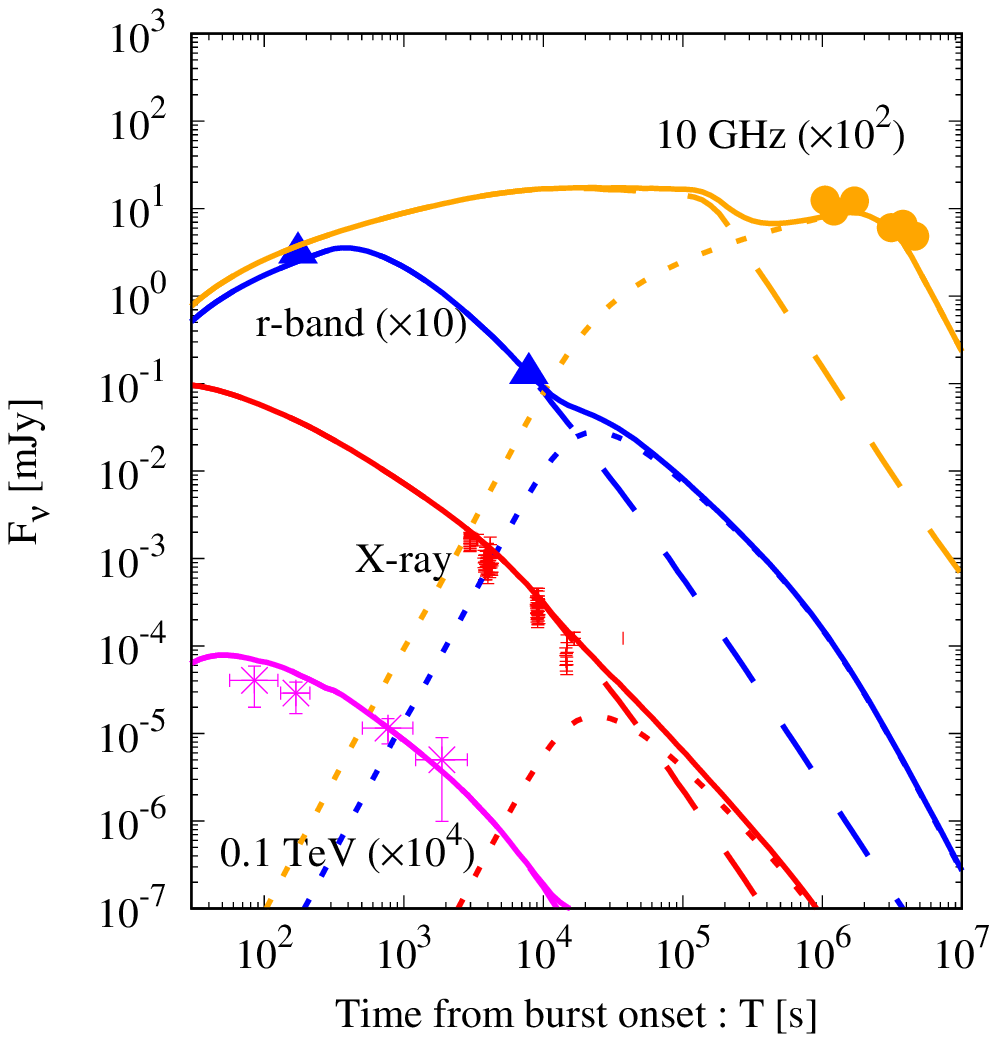}
\caption{
Multiwavelength light curves calculated by the two-component jet model for GRB~201216C -- solid lines are the sum of the narrow jet (dashed lines) and the wide jet (dotted lines).
The meanings of colors and observed data are the same as in Fig.~\ref{fig:GRB201216Cf}.
}
\label{fig:GRB201216C}
\end{figure}

Models for the afterglow of GRB~201216C are compared with observational data for the VHE (0.1~TeV), X-ray (5~keV), optical (r-band) and radio (10~GHz) bands in Figs.~\ref{fig:GRB201216Cf}, ~\ref{fig:GRB201216C_SED} and ~\ref{fig:GRB201216C}.
Taking the VHE data at 0.07--0.2~TeV
corrected for absorption by the extragalactic background light (EBL) in \citet{MAGIC2024},
we assume that the intrinsic photon index is $\sim2$ at all times and
convert the data to the flux density at 0.1~TeV.
The X-ray (5~keV) data is obtained from the {\it Swift} team 
website\footnote{https://www.swift.ac.uk/xrt\_curves/01013243/} \citep{Evans2007,Evans2009},
and the observed energy flux at 0.3-10 keV is converted to the flux density at 5~keV adopting a photon index of 1.67. 
The optical (r-band) and radio (10~GHz) data are extracted from \citet{Rhodes2022}.
Extinction in the r-band due to the Galaxy is assumed to be
$A_{r,\rm Gal} = 0.1$~mag, and that of
 the host to be $A_{r, \rm host} = 1.7$~mag \citep{Sato2023a}.

First, we consider the standard afterglow model of a single top-hat jet, taking a
typical opening half-angle of $\theta_0\sim0.1\,{\rm rad}$ \citep{Zhao2020}.
The radio afterglow is well explained when
the model parameters are chosen as follows:
$\theta_v =  0.0\,{\rm rad}$,
$\theta_0 =  0.1\,{\rm rad}$,
$E_{\rm{iso, K}} =  4.0 \times10^{53}\,{\rm erg}$,
$\Gamma_0 =  350$,
$n_0 =  1.0\,{\rm cm^{-3}}$,
$p =  2.3$,
$\epsilon_e =  0.38 $, 
$\epsilon_B =  1.0 \times 10^{-4}$ and
$f_e$ =  $0.2$.
We call this ``failed parameter set~I'' because the data at other wavelengths, e.g. the X-ray band, cannot be explained simultaneously.
Taking $T$ as the observer time,
as the typical frequency
$\nu_m\propto f_e^{-2} \epsilon_{e}^{2} \epsilon_{B}^{1/2} E_{\rm iso,K}^{1/2} T^{-3/2}$
crosses the observation bands, the flux reaches a maximum \citep{Sari1998}.
The cooling frequency is approximately given by
$\nu_c \propto \epsilon_B^{-3/2} E_{\rm iso,K}^{-1/2} n_0^{-1} T^{-1/2}$,
and the synchrotron-self-absorption frequency is estimated as
$\nu_a\propto f_e^{8/5} \epsilon_e^{-1} \epsilon_B^{1/5} E_{\rm iso,K}^{1/5} n_0^{3/5}$ for $\nu_a<\nu_m<\nu_c$
and 
$\nu_a\propto f_e^{3/5} \epsilon_B^{6/5} E_{\rm iso,K}^{7/10} n_0^{11/10} T^{-1/2}$ for $\nu_a<\nu_c<\nu_m$.
For the failed parameter set~I, the crossing time when $\nu_m$ intersects 10~GHz is $\simeq5\times10^{6}\,{\rm s}$, which is consistent with our numerical result to within a factor of two.
From $T\sim100$~s to $\sim3\times10^3$~s, 
frequencies $\nu_m$, $\nu_c$, $\nu_a$, 
$\nu_{\rm opt}=4.8\times10^{14}$~Hz (r-band) and $\nu_{\rm radio}=10$~GHz are ordered as $\nu_a < \nu_c < \nu_{\rm radio} < \nu_{\rm opt} < \nu_m$ (see the thick red line in Fig.~\ref{fig:GRB201216C_SED}(a)).
In this case, the optical and radio flux follows the scaling $F_{\nu}\propto T^{-1/4}$ \citep{Gao2013}.
Around $T\sim 4\times10^3$~s, the transition from the fast cooling phase ($\nu_m>\nu_c$) to the slow cooling phase ($\nu_m<\nu_c$) occurs.
After that, the break frequencies and the observation frequencies satisfy $\nu_a < \nu_{\rm radio} < \nu_{\rm opt} < \nu_m < \nu_c$ as shown in the thick blue line of Fig.~\ref{fig:GRB201216C_SED}(a), and the optical and radio flux follows $F_{\nu}\propto T^{1/2}$ \citep{Gao2013}.
The typical frequency $\nu_m$ crosses the r-band at $T\sim 1.5\times10^4$~s, 
after which the optical flux obeys the scaling $F_{\nu}\propto T^{-3(p-1)/4}\sim T^{-1}$ \citep{Gao2013}.
Subsequently, $\nu_m$ intersects 10~GHz around $T\sim 2\times10^6$~s (see the thick green line in Fig.~\ref{fig:GRB201216C_SED}(a)), and the 10~GHz flux decays in the same manner.
As shown in Fig.~\ref{fig:GRB201216Cf}(a), the predicted VHE, X-ray and optical flux exceed the observed values.

Next, we try to fit the VHE flux.
If $\epsilon_e$ is too small, then so is the Compton Y parameter, and
the VHE flux becomes too low \citep[e.g.][]{Sari2001,Nakar2009,Nava2021}.
When the cooling frequency $\nu_c$ is located between the optical and the X-ray bands, and $\nu_m$ is lower than the optical band,
both X-ray and optical flux are expected to follow $F_\nu\propto \epsilon_e^{p-1}$ \citep{Gao2013}.
For sufficiently small $\epsilon_e$ and $p>2$, the X-ray and optical afterglows 
also become too dim.
Hence, we set $\epsilon_e =  1.0 \times 10^{-2}$,
and the other parameters are kept the same as failed parameter set~I.
This parameter set is referred to as ``failed parameter set~II''.
The observed VHE gamma-ray light curve is well explained.
However, the radio afterglow is brighter than observed, and the X-ray and optical emission are still inconsistent with the observed data (red, blue and orange lines in Fig.~\ref{fig:GRB201216Cf}(b); see also the thick blue line in Fig.~\ref{fig:GRB201216C_SED}(b)).

As shown earlier, it is difficult to explain the observed multiwavelength afterglows at any time with a constant-density CBM in the single top-hat jet model.
\citet{MAGIC2024} proposed a wind-like ambient medium whose density decreases with radius $R$ as $R^{-2}$.
Then the optical, X-ray and VHE flux can be consistent with those observed, 
while the radio emission of the model still does not match the data. 
Another possibility is that $\epsilon_e$ or $\epsilon_B$ evolves with radius while $\theta_v = 0$.
However, then the X-ray afterglows may exhibit a shallow decay phase \citep{Fan2006,Granot2006,Ioka2006,Asano2024}, inconsistent with the observed monotonically decaying X-ray flux.
Hence, it is challenging for the single jet to describe well the observed multiwavelength afterglow emission from GRB~201216C.

Therefore, we introduce a two-component jet model.
Here we use the same parameters as in \citet{Sato2023a}, determined by comparing only with information from the GCN Circular by MAGIC, before the VHE light curves and spectra became available.
We have a ``narrow jet'' with
$\theta_v =  0.0\,{\rm rad}$,
$\theta_0 =  0.015\,{\rm rad}$,
$E_{\rm{iso, K}} =  4.0 \times10^{53}\,{\rm erg}$,
$\Gamma_0 =  350$,
$n_0 =  1.0\,{\rm cm^{-3}}$,
$p =  2.3$,
$\epsilon_e =  3.5 \times 10^{-2}$, 
$\epsilon_B =  6.0 \times 10^{-5}$
and
$f_e$ =  $0.4$ and
a ``wide jet'' with  
$\theta_v =  0.0\,{\rm rad}$,
$\theta_0 =  0.1\,{\rm rad}$,
$E_{\rm{iso, K}} =  1.0 \times 10^{53}\,{\rm erg}$,
$\Gamma_0 =  20$,
$n_0 =  1.0\,{\rm cm^{-3}}$,
$p =  2.8$,
$\epsilon_e =  0.1$, 
$\epsilon_B =  5.0 \times 10^{-5}$
and
$f_e =  0.2$.
Intriguingly, the VHE flux predicted by our model remains consistent with the observed data, even without modifying the parameters from \citet{Sato2023a}.

About $40\,{\rm s}$ after the burst trigger time, the SSC characteristic frequency, $2\gamma_m^2\nu_{m}$, where $\gamma_m$ is the minimum electron Lorentz factor, is expected to cross the VHE band,
at which the model VHE light curve reaches a maximum \citep[e.g.,][]{Nakar2009,Yamasaki2022}.
As the jet decelerates and the Doppler-boosted beaming cone becomes wider, progressively larger regions of the jet become visible to the observer until the edge of the top-hat jet is reached, after which the light curve steepens (``jet break'') \citep{Sato2021,Sato2023a,Sato2023b}.
Around the same time, physical lateral expansion of the jet may also occur and contribute to the jet break, although
relativistic hydrodynamical simulations show that the effect of the lateral expansion is not significant until the trans-relativistic ($\Gamma\lesssim10$) phase. 
The jet break time can be estimated by $T_{\rm jb}\sim 750~{\rm s}~(E_{\rm iso,K}/4\times10^{53}~{\rm erg})^{1/3}(n_0/1~{\rm cm^{-3}})^{-1/3}(\theta_0/0.015~{\rm rad})^{8/3}$ \citep{Sari1999} for the narrow jet.
Subsequently, another characteristic frequency, 
$2\gamma_m\gamma_0\nu_{m}$, intersects with 0.1~TeV at $\sim10^4\,{\rm s}$ (thin blue line in Fig.~\ref{fig:GRB201216C_SED}(c)),
where a critical electron Lorentz factor $\gamma_0$ is defined as $Y(\gamma_0)=1$ and $Y$ is the Compton Y parameter.
After this epoch, the decay index of the light curve changes \citep[e.g.,][]{Nakar2009,Yamasaki2022}.

The wide jet is contributing significantly only to the late time radio emission (see dotted lines in Fig.~\ref{fig:GRB201216C}).
Even in this case,
the wide jet parameters can be constrained 
from the observed radio data.
The radio flux reaches a maximum when $\nu_m$ crosses the relevant band.
The observed light curve at 10 GHz peaks at $T\sim2\times10^6\,{\rm s}$, 
requiring
$\nu_m\sim10\,{\rm GHz}$ at $T\sim2\times10^6\,{\rm s}$, 
before which $\nu_m >$10~GHz.
For $\nu_a<\nu<\nu_m$, the observed synchrotron flux density is approximately
$F_{\nu}(T)\propto \left(\frac{p-2}{p-1}\right)^{-\frac{2}{3}}\epsilon_{e}^{-\frac{2}{3}}\epsilon_{B}^{\frac{1}{3}}f_e^{\frac{5}{3}} E_{\rm iso,K}^{\frac{5}{6}} n_0^{\frac{1}{2}} T^{\frac{1}{2}}\nu^{\frac{1}{3}}$
\citep[e.g.][]{Sari1998,Gao2013},
which implies $\nu_a<10$~GHz until $T\sim10^7\,{\rm s}$.
In addition to $\nu_m(T\simeq2\times10^6\,{\rm s}) \simeq 10\,{\rm GHz}$, 
$F_{\nu=10\,{\rm GHz}}(T\simeq1\times10^6\,{\rm s})\simeq0.1\,{\rm mJy}$ as observed \citep{Rhodes2022} provides the two main conditions for constraining the parameters.
We adopt $p=2.8$ \citep{Sato2023a}
and take the number density $n_0$ to be the same as that determined for the narrow jet.
Following our suggestion in \citet{Sato2023a} that GRBs~180720B, 190114C, 190829A and 201216C have similar $E_{\rm iso,K}$, $\Gamma_0$ and $\theta_0$ for the narrow and wide jets, 
we set $E_{\rm iso,K}\simeq1.0\times10^{53}\,{\rm erg}$ \citep{Sato2023a}.
From the above two conditions
with $p=2.8$, $n_0=1.0\,{\rm cm^{-3}}$ and 
$E_{\rm iso,K}=1.0\times10^{53}\,{\rm erg}$,
we get
$\epsilon_{e} \simeq 1.2 f_e^{3/2}$ and
$\epsilon_{B} \simeq 1.1\times 10^{-6} f_e^{-2}$.
Therefore, our parameter values
$f_e\simeq0.2$, $\epsilon_e\simeq0.1$ and $\epsilon_B\simeq5\times10^{-5}$, 
which are almost consistent with the wide jet parameters reported in \citet{Sato2023a},
satisfy these conditions.


\subsection{GRB~221009A}
\label{subsec:GRB221009A}

\begin{figure*}
\centering
\begin{minipage}{0.31\linewidth}
\centering
\includegraphics[height=1.0\textwidth]{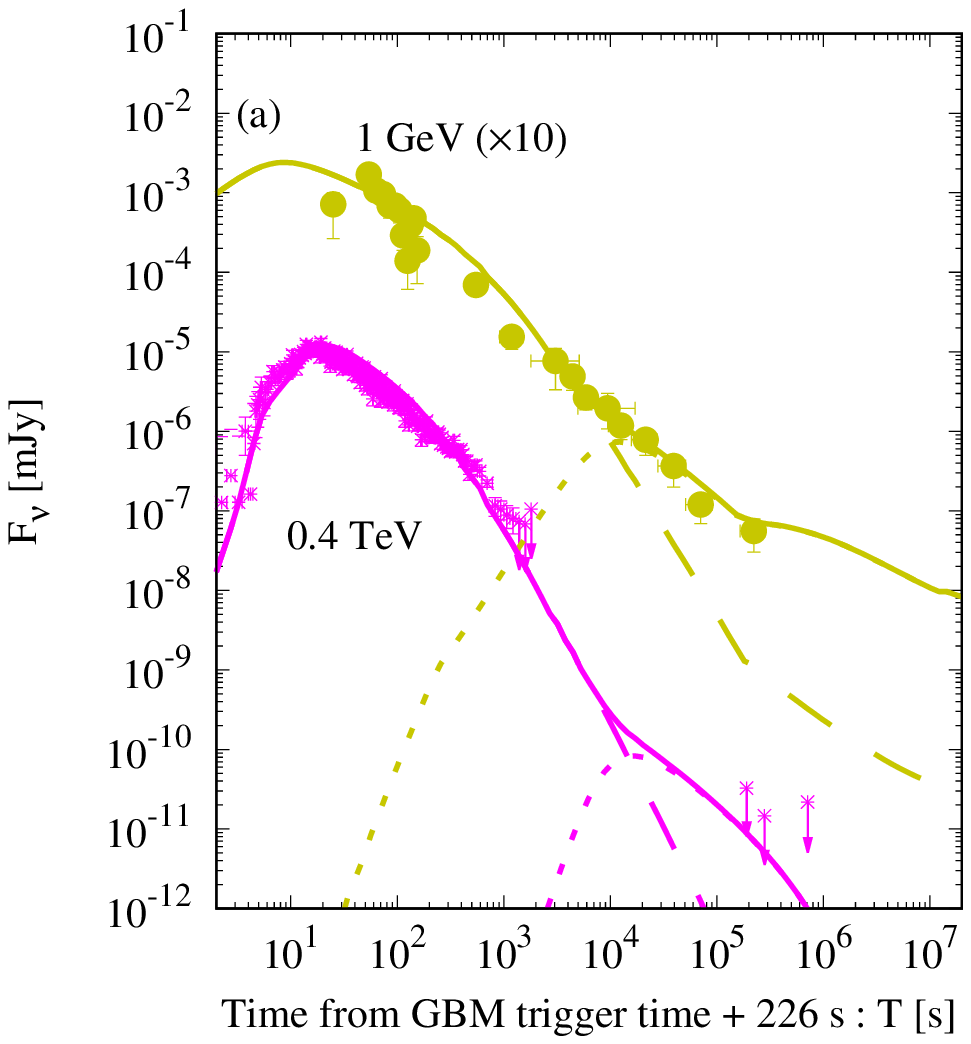}
\end{minipage}
\begin{minipage}{0.31\linewidth}
\centering
\includegraphics[height=1.0\textwidth]{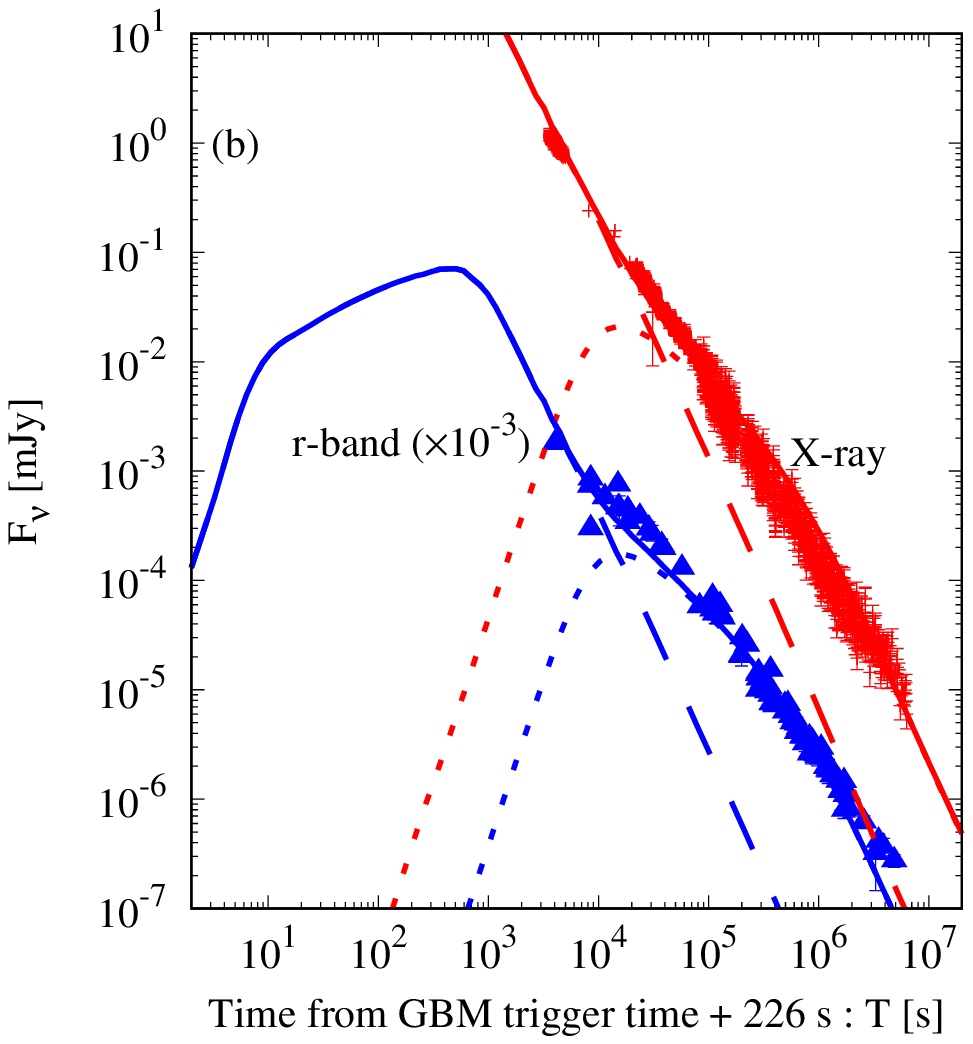}
\end{minipage}
\begin{minipage}{0.31\linewidth}
\centering
\includegraphics[height=1.0\textwidth]{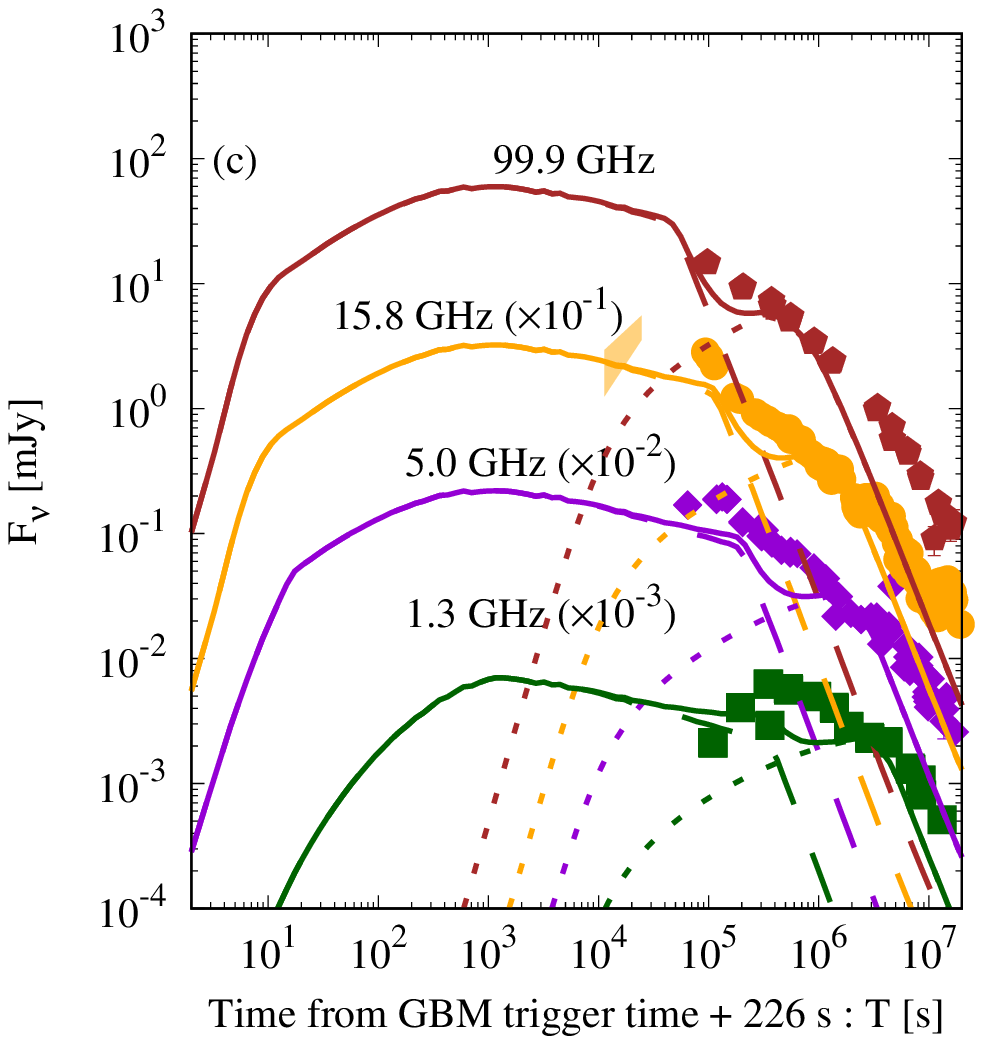}
\end{minipage}
\caption{
Observed data (VHE (0.4~TeV: magenta points), HE gamma-ray (1~GeV: yellow circles), X-ray (1~keV: red points), optical (r-band: blue upward triangles) and radio bands (1.3~GHz: green squares, 5.0~GHz: purple diamonds, 15.8~GHz: orange circles and 99.9~GHz: brown pentagons)) of GRB~221009A compared with multiwavelength afterglow emission from our two-component jet model -- solid lines are the sum of the narrow (dashed lines)
and wide (dotted lines) jets.
The upper limits of the VHE flux are shown with the magenta downward arrows.
The orange-shaded region will be dominated by the reverse-shock emission. 
In panel (a), we show the VHE (0.4~TeV: magenta) and HE gamma-ray (1~GeV: yellow) emission.
The panel (b) provides with the X-ray (1~keV: red) and optical (r-band: blue) afterglow light curves.
In panel (c), the radio afterglow emission (1.3~GHz: green, 5.0~GHz: purple, 15.8~GHz: orange and 99.9~GHz: brown) are given.
}
\label{fig:221009A}
\end{figure*}
\begin{figure}
\centering 
\includegraphics[width=0.35\textwidth]{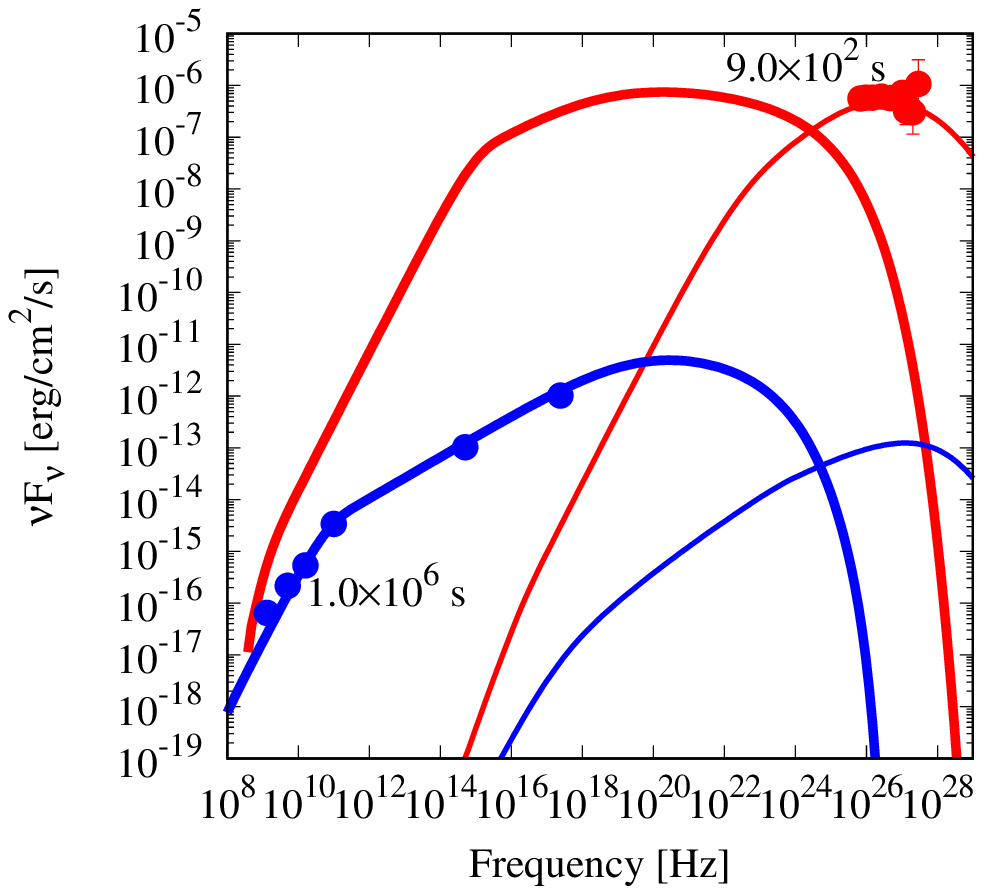}
\caption{
Model spectra of GRB~221009A for different observer times ($9.0\times10^2$~s since the {\it Fermi}/GBM trigger time plus 226~s in red and $1.0\times10^6$~s since the {\it Fermi}/GBM trigger time plus 226~s in blue), 
compared with observed data points.
Thick and thin solid curves show the synchrotron and SSC  spectra, respectively.
Note that the solid lines show the sum of the narrow and wide jets.
The narrow jet emission is dominated at $9.0\times10^2$~s since the {\it Fermi}/GBM trigger time plus 226~s (red lines), 
while the wide jet emission becomes non-negligible at $1.0\times10^6$~s since the {\it Fermi}/GBM trigger time plus 226~s (blue lines).
We use the unabsorbed optical data point in the r-band.
}
\label{fig:SED_09A}
\end{figure}

We compare the afterglow emission of GRB~221009A with observed data for VHE (0.4~TeV), high-energy (HE) (1~GeV), X-ray (1~keV), optical (r-band) and radio (1.3, 5.0, 15.8 and 99.9~GHz) bands in Figs.~\ref{fig:221009A} and \ref{fig:SED_09A}.
The VHE data and upper limits corrected for EBL attenuation are obtained from \citet{LHAASO2023a}, 
\citet{Mbarubucyeye2023} and \citet{HESS2023}.
The HE data is extracted from \citet{Tavani2023} and \citet{Axelsson2024}.
The X-ray data is derived from the {\it Swift} team website\footnote{https://www.swift.ac.uk/xrt\_curves/01126853/} and \citet{Williams2023}.
The optical data is taken from \citet{Fulton2023}, \citet{Kann2023} and \citet{Laskar2023}.
We assume extinction in the r-band for the Galactic and the host galaxy of $A_{r,\rm Gal}=4.1$~mag and $A_{r, \rm host}=0.7$~mag, respectively \citep{Blanchard2024}.
The radio data is taken from \citet{Bright2023}, \citet{Laskar2023}, \citet{OConnor2023} and \citet{Rhodes2024}.

The observed VHE flux exhibits the break where the light curve steepens at about 896~s after the {\it Fermi}/GBM trigger time plus 226~s \citep{LHAASO2023a},
while the observed X-ray ($\sim1-10$~keV) and optical light curves do not show such a steep decay at $\sim800$~s after the {\it Fermi}/GBM trigger time plus 226~s.
In order to explain these complicated multiwavelength afterglow behaviors, 
previous works have proposed the non-uniform density distribution, and they have supposed that a structured jet is also necessary \citep{Gill2023,BTZhang2023,Ren2024,Zheng2024}.
Therefore, even if the nontrivial CBM density profile is assumed, it is difficult for 
a single top-hat jet to describe the observed afterglows.
Another possibility is that the microphysical parameters evolve with the radius in a single top-hat jet model.
In this case, however, the predicted X-ray decay would be
shallower than observed \citep{Fan2006,Granot2006,Ioka2006,Asano2024}.
Hence, it is very challenging for the single uniform jet to well explain the broadband afterglow emission of GRB 221009A, as shown in several papers \citep{Gill2023,Sato2023b,BTZhang2023,Ren2024,Zheng2024}.

As noted in Section~\ref{sec:intro}, the detailed VHE light curve measured by LHAASO was not used in the modeling of \citet{Sato2023b}.
In our previous study, it was assumed that 
the observer time $T$ is set to be zero at the time of {\it Fermi}/GBM trigger.
By shifting the zero point of $T$ to 226~s after the {\it Fermi}/GBM trigger time,
the observed rising slope of the VHE light curve is almost consistent with the expected one for the constant CBM case \citep{LHAASO2023a}.

In order to describe the VHE flux, the model parameters determined by \citet{Sato2023b} need to be somewhat modified although the basic model itself remains the same. 
For the narrow jet, we set
$\theta_v=0\,{\rm rad}$, 
$\theta_0=1.8\times10^{-3}\,{\rm rad}$,
$E_{\rm iso,K}=3.0\times10^{55}\,{\rm erg}$, 
$\Gamma_0=800$,
$n_0 =0.1\,{\rm cm^{-3}}$, 
$p=2.2$, 
$\epsilon_e=2.0\times10^{-2}$,
$\epsilon_B=1.0\times10^{-4}$
and $f_e=0.2$.
Then, the rising part and peak emission in the VHE band can be explained.
The light curves have peaks at the transition from the coasting phase to the self-similar adiabatic expansion phase \citep{Sari1997}.
The deceleration time is given by 
$T_{\rm dec} \sim 16~{\rm s}~(E_{\rm iso, K}/3.0\times10^{55}\,{\rm erg})^{1/3}(n_0/0.1\,{\rm cm^{-3}})^{-1/3} (\Gamma_0/800)^{-8/3}$ \citep{Sari1997},
which is almost consistent with the observed VHE gamma-ray 
light curve.
Although the initial jet opening half-angle is smaller than that adopted in our previous work \citep{Sato2023b}, it still satisfies $\theta_0>\Gamma_0^{-1}$.

Our narrow jet has a large bulk Lorentz factor, so here
we examine the relationship between the radius of the prompt emission (i.e., internal shock) 
and the deceleration radius for the narrow jet.
The former is given by
$R_{\rm IS}\sim 2c\delta t \Gamma_0^{2}$, where $\delta t$ is the variability timescale of the prompt emission
\citep[e.g.][]{Piran2004,Zhang2019}.
When we use $\delta t\sim10\,{\rm ms}$ \citep{Lesage2023} and our narrow jet parameters, 
we get $R_{\rm IS}\sim1\times10^{15}\,{\rm cm}$.
On the other hand, the deceleration radius of the external shock for our narrow jet is estimated to be 
$R_{\rm dec}\sim 4\times10^{17}~{\rm cm}~(E_{\rm iso, K}/3\times10^{55}~{\rm erg})^{1/3}(n_0/0.1~{\rm cm^{-3}})^{-1/3}(\Gamma_0/800)^{-2/3}$ \citep[e.g.][]{Piran2004,Zhang2019}.
Therefore, we 
have
$R_{\rm IS}<R_{\rm dec}$, 
which is a standard order of characteristic radii for typical GRBs.
Moreover, using a model of \citet{Matsumoto2019}, 
we get small optical depth for electron-positron pair production by the highest energy prompt photons of $\sim100$~GeV.
In other words, the minimum bulk Lorentz factor of the jet of about 300 is derived from the observed prompt emission properties \citep{Frederiks2023,Gao2023}, which is consistent with our value of $\Gamma_0$ for the narrow jet.

The late optical, X-ray and HE gamma-ray emission from our narrow jet decay more rapidly than observed due to the jet break effect (dashed lines in Figs.~\ref{fig:221009A}(a) and (b)).
Therefore, the wide jet is necessary in addition to the narrow jet.
In order to explain the observed late optical, X-ray and HE gamma-ray flux, 
we adopt the model parameters of the wide jet as
$\theta_v=0\,{\rm rad}$, 
$\theta_0=6.0\times10^{-2}\,{\rm rad}$,
$E_{\rm iso,K}=3.0\times10^{53}\,{\rm erg}$, 
$\Gamma_0=30$,
$n_0 =0.1\,{\rm cm^{-3}}$, 
$p=2.2$,
$\epsilon_e=0.1$, 
$\epsilon_B=1.5\times10^{-5}$
and $f_e=0.1$.
The electron spectral index is consistent with the observed X-ray photon index of $1.8$.
The observed optical, X-ray and HE gamma-ray afterglows at the late time are consistent with the emission from the wide jet (see the thick blue line in Fig.~\ref{fig:SED_09A}).
After the peak in the optical and X-ray light curves, the wide jet enters the self-similar adiabatic expansion phase.
Then, the typical frequency $\nu_m$ is below the r-band frequency, and the cooling frequency $\nu_c$ is between the optical and X-ray bands.
Therefore, the optical and X-ray light curves follow $F_\nu\propto T^{(2-3p)/4}\sim T^{-1.2}$ and $F_\nu\propto T^{3(1-p)/4}\sim T^{-1.0}$,
respectively \citep{Gao2013}, which are almost consistent with the observed ones (blue-solid and red-solid lines in Fig.~\ref{fig:221009A}(b)).
Around $T\sim8\times10^4$~s, the shallow break in the observed X-ray light curve is observed \citep{Williams2023}, which may indicate a transition from the narrow jet-domineted to the wide jet-dominated phases \citep{Sato2023a}.

The SSC emission from our narrow jet can almost reach $\approx 10$~TeV (see the thin red line in Fig.~\ref{fig:SED_09A}).
A large Lorentz factor allows photons that are not significantly affected by Klein-Nishina suppression in the comoving frame to reach an observed photon energy of $\approx10$~TeV.
On the other hand, the observed VHE gamma-ray energy spectrum is still not well described by the SSC radiation. Although may be within the uncertainty range of the EBL model, an additional component such as hadronic emission might be necessary \citep{BTZhang2023}.

Our radio flux sometimes overpredicts the observed data, while the differences are within a factor of three. 
This may be attributed to the uncertainty of our simple model. 
If the absorption frequency $\nu_a$, the typical frequency $\nu_m$, 
and the radio band frequency $\nu_{\rm radio}\sim1$--10~GHz satisfy
$\nu_a<\nu_{\rm radio}<\nu_m$,
then the radio flux scales as $F_\nu\propto\epsilon_e^{-2/3}\epsilon_B^{1/3}$.
On the other hand, the Compton Y parameter 
affected by the Klein-Nishina effect can be written by 
$Y \sim10^2~(\epsilon_e/0.1)^{1/2}(\epsilon_B/10^{-5})^{-1/2}$
in the slow cooling regime \citep[e.g.,][]{Nakar2009,Yamasaki2022}.
Parameter combinations such as large $\epsilon_e$ and/or small $\epsilon_B$ 
(e.g., $\epsilon_e \sim 0.1$, $\epsilon_B \sim 10^{-5}$) can suppress the radio flux while enhancing the VHE emission through large value of $Y$.
Although the late-time radio data are insufficient to tightly constrain the microphysical parameters,
the VHE observation may provide stronger constraints on $\epsilon_e$ and $\epsilon_B$.

As shown in Fig.~\ref{fig:221009A}(c), our radio light curves sometimes do not explain observed ones.
In particular, the radio emission at $T\sim 2\times10^4$~s of this model is inconsistent with the observed bump in the 15.8~GHz band (see the orange shaded region in Fig.~\ref{fig:221009A}(c)).
The reverse-shock emission potentially contributes to this radio bump \citep{Gill2023,OConnor2023,BTZhang2023,Ren2024,Rhodes2024,Zheng2024}.
In general, outflows with large isotropic-equivalent kinetic energy
tend to have large radius of reverse shock crossing \citep{Kobayashi1999}.
In the case of our narrow jet, however, the jet has a large bulk Lorentz factor, 
leading to earlier reverse-shock crossing and more rapid flux decay than observed.
Below, using our wide jet parameters, we estimate 
the reverse-shock emission
in the thin shell case to show that it is roughly consistent with observation.
The observed reverse shock flux density in the thin shell case after the reverse shock crossing can be expressed by 
$F_{{\rm r1},\nu}(T) \propto \left(\frac{p_{\rm r}-2}{p_{\rm r}-1}\right)^{-\frac{7}{3}} \Gamma_0^{-\frac{101}{105}} f_{e,{\rm r}}^{-1} \epsilon_{e,{\rm r}} E_{\rm iso,K}^{\frac{88}{105}} {n_0}^{-\frac{41}{35}} T^{\frac{18}{35}} \nu^{2}$ 
for $\nu<\nu_{a,{\rm r}}$ and
$F_{{\rm r2},\nu}(T)\propto \left(\frac{p_{\rm r}-2}{p_{\rm r}-1}\right)^{-\frac{2}{3}}\Gamma_0^{-\frac{31}{35}} f_{e,{\rm r}}^{\frac{5}{3}} \epsilon_{e,{\rm r}}^{-\frac{2}{3}} \epsilon_{B,{\rm r}}^{\frac{1}{3}} E_{\rm iso,K}^{\frac{121}{105}} n_0^{\frac{2}{15}} T^{-\frac{16}{35}} \nu^{\frac{1}{3}}$
for $\nu_{a,{\rm r}}<\nu<\nu_{m,{\rm r}}$ \citep[e.g.][]{Gao2013,Resmi2016}.
When we fix the same values of $E_{\rm iso,K}$, $\Gamma_0$ and $n_0$
as those of the wide jet,
the expected reverse shock flux at $\nu=15.8$~GHz is estimated as
$F_{{\rm r1},\nu} \sim 44~{\rm mJy} \left(\frac{T}{2\times10^4~{\rm s}}\right)^{\frac{18}{35}}$ 
and
$F_{{\rm r2},\nu}\sim 44~{\rm mJy} \left(\frac{T}{2\times10^4~{\rm s}}\right)^{-\frac{16}{35}}$,
where we adopt $f_{e,{\rm r}}=0.1$, $\epsilon_{e,{\rm r}}=0.065$, $\epsilon_{B,{\rm r}}=0.4$ and $p_{r}=2.3$.
The light curve has a peak
when the absorption frequency crosses 15.8~GHz, and the peak time is estimated as
$T\sim 2\times10^{4}\left(\frac{f_{e,{\rm r}}}{0.1}\right)^{\frac{140}{51}} \left(\frac{\epsilon_{e,{\rm r}}}{0.065}\right)^{-\frac{175}{102}}\left(\frac{\epsilon_{B,{\rm r}}}{0.4}\right)^{\frac{35}{102}} \left(\frac{\nu}{15.8~{\rm GHz}}\right)^{-\frac{175}{102}}\,{\rm s}$
\citep{Gao2013,Resmi2016}.
These expected peak flux and peak time are roughly consistent with the observed radio emission around $T\sim2\times10^4$~s.
In this paper, we focus on the early VHE gamma-ray afterglows, 
and then the reverse shock emission is less significant.
More detailed study of the radio emission remains to be future work.

In GRB~221009A, LHAASO observed the early break in the VHE light curve \citep{LHAASO2023a}, which may be originated from a jet break of the narrow jet \citep{LHAASO2023a,Ren2024,Zheng2024}.
Therefore, only for GRB~221009A, the emission from the narrow jet is strongly suggested \citep{LHAASO2023a,BTZhang2023,Ren2024,Zheng2024}.
The upper limits in~the~VHE~bands at the late epoch were reported \citep{HAWC2022,HESS2023}.
Due to the jet break effect, it is hard for the narrow jet to explain the late ($T\gtrsim7\times10^3\,{\rm s}$) afterglows.
A baryon-loaded jet ($\Gamma_0\ll 100$) is necessary to fit the late VHE gamma-ray afterglows as well as the X-ray and optical emission.
The baryon-loaded jet such as our wide jet can also radiate the SSC emission at the late time.


\section{Discussion}
\label{sec:discussion}

\begin{table*}
\centering
\caption{Summary of parameters adopted in our two-component jet model and our prompt efficiency $\eta$.}
\label{tab:GRB_model}
\begin{tabular}{lcccccccccc}
\hline
&  ~$\theta_v$~[rad]~ & $\theta_0$~[rad] & ~$\Gamma_0$~ & $E_{\rm{iso, K}}$~[erg]  & ~$n_0$~${[\rm cm^{-3}]}$~ & ~$p$~ & $\epsilon_e$ & ~$\epsilon_B$~ & ~$f_e$~ & ~$\eta$~  \\
\hline 
GRB~180720B${}^{\rm a}$ &&&&&&&&&\\
narrow jet & \multirow{2}{*}{~$0.0$~} & ~$0.015$~ & ~$350$~ & $4.0\times10^{53}$ & \multirow{2}{*}{~$10$~} & ~$2.4$~  & ~$ 5.0\times10^{-3}$~ & $5.0\times10^{-4}$ & ~$0.2$~ & ~$0.6$~ \\
wide jet &  & ~$0.1$~ & ~$20$~ & $1.0\times10^{53}$ & & ~$2.2$~  & ~$9.0\times10^{-2}$~ & $9.0\times10^{-4}$ & ~$0.4$~ & ~$-{}^{\rm c}$~ \\
\hline
GRB~190114C${}^{\rm a}$  &&&&&&&&&\\
narrow jet & \multirow{2}{*}{~$0.0$~} & ~$0.015$~ & ~$350$~ & $4.0\times10^{53}$ & \multirow{2}{*}{~$3.0$~} & ~$2.8$~  & ~$ 9.0\times10^{-3}$~ & $6.0\times10^{-5}$ & ~$0.1$~ & ~$0.4$~ \\
wide jet &  & ~$0.1$~ & ~$20$~ & $1.0\times10^{53}$ & & ~$2.6$~  & ~$8.0\times10^{-2}$~ & $9.0\times10^{-4}$ & ~$0.2$~ & ~$-{}^{\rm c}$~ \\
\hline
GRB~190829A${}^{\rm a}$ &&&&&&&&&\\
narrow jet & \multirow{2}{*}{~$0.031$~} & ~$0.015$~ & ~$350$~ & $4.0\times10^{53}$ & \multirow{2}{*}{~$0.01$~} & ~$2.44$~  & ~$3.5\times10^{-2}$~ & $6.0\times10^{-5}$ & ~$0.2$~ & ~$0.4{}^{\rm d}$~ \\
wide jet &  & ~$0.1$~ & ~$20$~ & $1.0\times10^{53}$ & & ~$2.2$~  & ~$0.29$~ & $1.0\times10^{-5}$ & ~$0.35$~ & ~$-{}^{\rm c}$~ \\
\hline
GRB~201216C${}^{\rm a}$ &&&&&&&&&\\
narrow jet & \multirow{2}{*}{~$0.0$~} & ~$0.015$~ & ~$350$~ & $4.0\times10^{53}$ & \multirow{2}{*}{~$1.0$~} & ~$2.3$~  & ~$3.5\times10^{-2}$~ & $6.0\times10^{-5}$ & ~$0.4$~ & ~$0.5$~ \\ 
wide jet &  & ~$0.1$~ & ~$20$~ & $1.0\times10^{53}$ & & ~$2.8$~  & ~$0.1$~ & $5.0\times10^{-5}$ & ~$0.2$~ & ~$-{}^{\rm c}$~ \\
\hline
GRB~221009A${}^{\rm b}$ &&&&&&&&&\\
narrow jet & \multirow{2}{*}{~$0.0$~} & ~$0.0018$~ & ~$800$~ & $3.0\times10^{55}$ & \multirow{2}{*}{~$0.1$~} & ~$2.2$~  &  ~$2.0\times10^{-2}$~ & $1.0\times10^{-4}$ & ~$0.2$~ & ~$0.3$~ \\
wide jet &   & ~$0.06$~ & ~$30$~ & $3.0\times10^{53}$ & & ~$2.2$~ & ~$0.1$~ & $1.5\times10^{-5}$  & ~$0.1$~ & ~$-{}^{\rm c}$~ \\
\hline
\end{tabular}
\begin{tablenotes}
\item {\bf Notes.} \\
${}^{\rm a}$~\citet{Sato2023a}.\\
${}^{\rm b}$~This work.\\
${}^{\rm c}$~N/A (The prompt emission from our wide jet is negligible because of small values of $\Gamma_0$).\\
${}^{\rm d}$~We use on-axis $E_{\rm \gamma,iso}(\theta_v=0)$ (see \citet{Sato2021}).\\
\end{tablenotes}
\end{table*}

We compare our two-component jet model for GRB~221009A 
with other previous models by other authors 
(\S~\ref{subsec:compara}), and 
summarize similarities (\S~\ref{subsec:common})
and differences (\S~\ref{subsec:difference}) 
in our results for five VHE-detected events as 
inferred from the two-component jet model 
(see Table~\ref{tab:GRB_model} for summary).
Finally, we argue physical implications and future prospects for the model
in \S~\ref{subsec:implication}.

\subsection{Comparison with other structured jet models for
GRB~221009A}
\label{subsec:compara}

GRB~221009A shows the early break in the VHE band
about 896~s after the {\it Fermi}/GBM trigger time plus 226~s, 
implying a highly collimated jet with an initial opening angle $\theta_0\approx2\times10^{-3}$~rad.
Note that the X-ray ($\sim1-10$~keV) and optical emission 
around this break time
was not detected.
Previous studies have modeled its multiwavelength afterglow 
using various types of non-uniform, structured jets,
in which the isotropic-equivalent kinetic energy varies as a function of polar angle
\citep[e.g.,][]{BTZhang2023,Ren2024,Zheng2024}.
Despite differences in the functional forms of their energy distributions,
these models consistently feature a narrow, high-energy core.
The core sizes in such models are typically comparable to, but slightly larger than,
the opening angle assumed in our narrow top-hat jet model.
Moreover, the isotropic-equivalent kinetic energy of these cores,
on the order of $\sim 10^{55}$~erg,
is similar to that of our jet.
Taken together, these results --- 
both from previous structured jet models and our own ---
support the existence of
extremely narrow ($\sim 10^{-3}$~rad) 
and energetic ($E_{\rm K,iso}\sim 10^{55}$~erg) jets in GRB~221009A.
On the other hand, the ambient density and microphysical parameters in our model are significantly different between our work and previous studies \citep{BTZhang2023,Ren2024,Zheng2024}.

Previous works have assumed the non-uniform density distribution of the external medium \citep{BTZhang2023,Ren2024,Zheng2024}.
In contrast, in this study, we have examined the 
uniform
CBM case.
At present, both the constant CBM and the wind-like medium scenarios remain possible.
Further investigations would be useful for probing the environment of this event.

The radio emission from our two-component jet does not describe the observed radio data well.
In this work, we only consider the forward-shock emission.
On the other hand, the previous studies have also calculated the reverse-shock emission \citep{BTZhang2023,Ren2024,Rhodes2024,Zheng2024}.
However, it is still challenging to satisfactorily explain the observed radio emission even when the reverse shock component is included \citep{Ren2024,Rhodes2024,Zheng2024}.
This may imply the existence of the additional component
that only emits the radio afterglow.

\subsection{Similarities among VHE GRBs}
\label{subsec:common}

For VHE events, luminosities in the X-ray and optical bands are much brighter than those of typical long GRBs \citep{Ror2023},
while their radio luminosities tend to be dim \citep{Rhodes2020,Laskar2023}.
When the typical initial jet opening half-angle of $\theta_0\sim0.1\,{\rm rad}$ is introduced to describe the X-ray and optical afterglows, the radio emission becomes bright \citep{Sato2021}.
If the typical frequency $\nu_m$ is above the radio bands and $\nu_a$ is below them, then
the observed dim radio afterglows requires large $\epsilon_e$ and/or small $\epsilon_B$
\citep{Sato2021,Sato2023a,Sato2023b},
as already discussed in \S~\ref{subsec:GRB221009A}.
In this case, however, the VHE emission would 
become brighter than observed due to the large 
Compton~Y parameter \citep[e.g.][]{Sari2001,Nakar2009,Nava2021}.
Here, the small initial jet opening half-angle is adopted to give
dim radio afterglows.
Yet, this also reduces the X-ray and optical fluxes, making it difficult to explain the observed brightness in those bands.
Therefore, it is challenging for a single jet to describe the multiwavelength afterglows, from radio to VHE bands, simultaneously.
The VHE events may be among essentially different
kinds from ordinary GRBs, and
they need to have a two-component jet.

The isotropic kinetic energies ($E_{\rm iso,K}$) of the narrow and wide jets for GRB~201216C are 
comparable to those of our previously modeled two-component jets 
for other VHE GRBs~180720B, 190114C and 190829A \citep{Sato2023a}.
In contrast,  $E_{\rm iso,K}$ 
of the narrow jet for GRB~221009A is 
approximately
two orders of magnitude larger than that of these VHE events.
The collimation-corrected kinetic energies of our narrow and wide jets for five VHE GRBs are estimated to be $E_{\rm jet,K}\sim5\times10^{49}\,{\rm erg}$ and $\sim5\times10^{50}\,{\rm erg}$, respectively.
All VHE GRBs possess similar values of $E_{\rm jet,K}$.
Although VHE GRBs tend to exhibit large $E_{\rm iso,K}$
\citep[e.g.][]{Yamazaki2020},
their collimation-corrected kinetic and prompt gamma-ray energies can remain within typical ranges.
Notably, 
the collimation-corrected kinetic energy $E_{\rm iso,K}$ of the wide jet is greater than that of the narrow jet, which is consistent with results from hydrodynamic simulations \citep[e.g.][]{Tchekhovskoy2008}.

It is an observational fact that among five VHE gamma-ray events so far, four of them have large isotropic-equivalent gamma-ray energy \citep[e.g.][]{Yamazaki2020}: $E_{\rm iso,\gamma}\sim6.0\times10^{53}\,{\rm erg}$ for GRB~180720B \citep{HESS2019}, $E_{\rm iso,\gamma}\sim3.0\times10^{53}\,{\rm erg}$ for GRB~190114C \citep{MAGIC2019a}, $E_{\rm iso,\gamma}\sim6.2\times10^{53}\,{\rm erg}$ for GRB~201216C and $E_{\rm iso,\gamma}\sim1.0\times10^{55}\,{\rm erg}$ for GRB~221009A \citep{Burns2023,Frederiks2023}.
One exception was a low-luminosity event, GRB~190829A, that had
$E_{\rm iso,\gamma}\sim3.2\times10^{49}$~erg \citep{Chand2020}.
Our narrow jet for GRB~221009A has a large isotropic kinetic energy,
$E_{\rm iso,K}=3.0\times10^{55}\,{\rm erg}$, and its prompt efficiency is $\eta\sim E_{{\rm iso}, \gamma}/(E_{\rm iso, \gamma}+E_{\rm iso, K})\sim0.3$.
The narrow jets of the other VHE GRBs have a similar value of $E_{\rm iso,K}$ ($\sim4.0\times10^{53}\,{\rm erg}$: see Table~\ref{tab:GRB_model}).
Then, the efficiency of the narrow jet for GRBs~180720B, 190114C, 190829A and 201216C is $\eta\sim0.6$, $\sim0.4$, $\sim0.4$ and $\sim0.5$, respectively, where we assume that GRB~190829A has the value of $E_{\rm iso,\gamma}(\theta_v = 0.0\,{\rm rad})\sim2.7\times10^{53}\,{\rm erg}$ based on our off-axis jet model \citep{Sato2021}.
All VHE gamma-ray events have similar efficiency of the prompt emission if they are viewed on-axis.

In our modeling, VHE GRBs have $\epsilon_e\sim0.01-0.1$ and $\epsilon_B\sim10^{-5}-10^{-3}$.
The distribution of $\epsilon_e$ for long GRBs without VHE gamma-ray detections is roughly similar to that of GRBs with VHE gamma-ray detection.
On the other hand, the distribution of $\epsilon_B$ for non-VHE GRBs has somewhat broader than that of VHE gamma-ray events, 
and it has the upper end at $\epsilon_B\sim10^{-2}$
\citep{Santana2014}, so the VHE GRBs tend to have smaller values of $\epsilon_B$ than events without VHE gamma-ray detection.
At present, the VHE gamma-ray photons can only be detected under suitable observing conditions \citep{MAGIC2019a}. It may not be necessary that there is any difference between bursts with and without VHE gamma-ray detection in the intrinsic properties of GRBs.

\subsection{Differences among VHE GRBs}
\label{subsec:difference}

As shown in Table~\ref{tab:GRB_model},
VHE GRBs occur in the ambient medium with various densities,
$n_0=0.01$--10~cm$^{-3}$,
which is broadly consistent with typical long GRBs 
\citep[e.g.,][]{KZhang2021,Chrimes2022}.
Hence, the ambient number density may hardly affect the detectability of the VHE emission.
In our modeling, some VHE GRBs have small ambient density, $n_0\lesssim0.1$~cm$^{-3}$.
If the GRB jets propagate into (super)bubbles, 
the CBM could have small density.
As a typical example, let us consider a wind from progenitor
Wolf-Rayet stars. If the wind with a velocity of $v_w=v_8\times10^8~{\rm cm~s^{-1}}$ makes a spherical cavity with the radius  $r_c=r_{19}\times10^{19}$~cm, then
the number density just inside the cavity wall is roughly 
given by $\sim 4\times10^{-4}\,{\rm cm^{-3}}
v_8^{-2}n_0T_{\rm eV}$, where 
the interstellar medium just outside of the cavity
has the number density $n_0$ and temperature
$T_{\rm eV}$ in units of eV
\citep{Weaver1977,Chrimes2022}.
Note that some hydrodynamics simulations of the wind cavity 
have shown that the stalled wind has an almost
flat density profile, and that its density is about 3 to 5 orders of magnitude smaller than the surrounding (unshocked) interstellar medium density \citep{Chrimes2022,Dwarkadas2022}.
It may be possible that GRBs occur in such a low-density stalled wind region, and in this case, we can approximate that the GRB arises in the constant rarefied ambient medium as assumed in our study.
Furthermore, global magnetohydrodynamics simulations of the disk and halo of a star forming galaxy that solve dynamical evolution of the interstellar medium have shown that the galaxy is occupied by non-negligible volume of the hot ($\gtrsim10^5$~K), rarefied ($\lesssim10^{-2}$~cm$^{-3}$) medium \citep{deAvillez2005}.
It may also be possible that GRBs arise in such a low-density constant medium.
Moreover, for GRB~190829A, very-long baseline interferometry observations provided a constraint on the $E_{\rm iso,K}/n_0$ ratio, suggesting a low ambient number density \citep{Salafia2022}.

VHE GRBs have various radio luminosities \citep{Rhodes2020,Rhodes2022,Laskar2023}.
The radio luminosities of GRBs~190829A and 221009A are about an order of magnitude dimmer than those of other VHE gamma-ray events \citep{Rhodes2020,Rhodes2022,Laskar2023}.
The densities of the CBM of GRBs~190829A and 221009A are also smaller than those of other VHE gamma-ray events (see Table~\ref{tab:GRB_model}).
Long GRBs may be categorized into two types, radio-loud and radio-quiet GRBs \citep{KZhang2021,Lloyd-Ronning2022,Chakraborty2023}.
Radio-quiet bursts may tend to have a small $n_0$ \citep{Sato2021}.
Note that afterglows at other frequencies are also affected by the ambient number density, but other model parameters could contribute significantly to flux in other bands.
For GRB~221009A, the reverse-shock emission may be dominated in the radio bands \citep{Bright2023,Laskar2023,OConnor2023,BTZhang2023,Ren2024,Zheng2024}, so it is not simple to describe the radio emission.
Further studies of radio afterglows are left for future work.

\subsection{Implications of two-component jets}
\label{subsec:implication}

The fraction of long GRBs with the X-ray shallow decay phase without VHE emission is about 88\% \citep{Yamazaki2020},
while all VHE GRBs have a less noticeable shallow decay phase \citep{Fraija2019b,Yamazaki2020,Ror2023}.
If the structured jets are viewed off-axis, the shallow decay phase will appear in the X-ray light curve \citep{Beniamini2020a,Beniamini2020b}.
Hence, the emission from around the jet core may be observed 
in VHE gamma-ray events (i.e., $\theta_v\approx0$).
\citet{Sato2021,Sato2023a} supposed that VHE GRB~190829A may be viewed off-axis, although the X-ray afterglow of GRB~190829A has no shallow decay phase \citep{Yamazaki2020}.
When the jet core angle is small, the shallow decay phase tends to look less noticeable \citep{Oganesyan2020}.
VHE GRBs may have a small jet core angle like our narrow jet, or they may be seen off-axis with a much smaller viewing angle.
Several models have been proposed to explain the shallow decay phase (e.g., energy injection model \citep{Granot2006b,Nousek2006,Zhang2006,Kobayashi2007}). 
The differences between the models can be seen in the evolution of the broadband spectrum in the VHE gamma-ray regime \citep[e.g.][]{Murase2010,Murase2011,Asano2024}.
Therefore, the VHE emission could be used for the discrimination among various models.

Here, we consider the case where the narrow jet is viewed off-axis, while the wide jet is seen on-beam.
Note that the opening half-angle of the wide jet is larger than the viewing angle.
Then, the X-ray luminosity takes maximum by the wide jet at $T\sim10^5\,{\rm s}$.
So far, such X-ray afterglows have not been observed. 
The wide jet may not radiate the prompt emission due to its large optical depth.
On the other hand, the prompt emission from the narrow jet becomes dim and soft due to the relativistic beaming effect,
and they have not been detected by previous and current detectors. 
Subsequent follow-up observations have not yet been performed for such events because of the faint prompt emission.
However, in the near future, faint soft X-ray transients from the narrow jet viewed off-axis may be detected by the Einstein Probe 
\citep[e.g.,][]{Gao2024,Yuan2025}, which will give us new insights into the jet structure.

In the off-axis viewing case, low-luminosity GRBs can be detected \citep[e.g.][]{Yamazaki2003,Ramirez-Ruiz2005,Sato2021}.
VHE gamma-ray photons with a significance of $\sim3\,\sigma$ from GRB~201015A were reported by Major Atmospheric Gamma Imaging Cherenkov (MAGIC) telescopes, 
and this event is classified as a low-luminosity GRB \citep{Blanch2020}.
Such observed features may be interpreted as evidence for an off-axis jet scenario.
In the future, CTA 
may increase the number of low-luminosity GRBs detected at the $\sim3\sigma$ level in the VHE regime.

The microphysical parameters of the narrow and wide jets have different values in our modeling.  
If the microphysical parameters evolve slowly with the radius, the microphysical parameters of both jets could have different values, since the dominant emission from each jet originates at different radii.
In our model, the characteristic emission radii of the narrow and wide jets are $\sim10^{17}\,{\rm cm}$ and $\sim10^{18}\,{\rm cm}$, respectively.
For example, when $\epsilon_e$ follows as $\epsilon_e=0.01(r/10^{17}\,{\rm cm})$, we get $\epsilon_e\sim0.01$ for the narrow jet and $\epsilon_e\sim0.1$ for the wide jet, which are roughly consistent with our modeling results (see Table~\ref{tab:GRB_model}).
In addition, the differences of the jets and/or the CBM properties may cause variation of 
the microphysical parameters between the narrow  and the wide jets \citep{Sato2023a}.
Further investigation into relativistic shock physics are necessary.

A jet with a small initial opening half-angle is necessary to explain the observed multiwavelength afterglow data of VHE GRBs.
Such a narrow jet can be produced by a Poynting-flux-dominated jet \citep{Zhang2024}.
Numerical simulations will be essential to study the formation mechanism of such narrow jets.

For the single top-hat jet, the time evolution of a polarization degree shows two peaks due to the jet break effect \citep{Shimoda2021}.
If the GRB jet consists of two components,
four peaks may appear in the time variation of the polarization degree 
in the optical and radio bands.
The existence of the two-component jet could be indicated by the polarimetric observations.
In addition, dips and/or bumps in light curves may also indicate the existence of the two-component jet \citep{Sato2023a}.
The temporal index of the light curve would change 
around the transition epoch 
from the narrow jet-dominated to the wide jet-dominated phases.
Therefore, long-term observations are crucial to distinguish between the two components.
In our modeling, VHE GRBs tend to have similar values for the collimation-corrected kinetic energy, the Lorentz factor and the jet opening angle for each jet component (see Table~\ref{tab:GRB_model}).
In the future, quasi-simultaneous optical observations with, e.g., the Vera C. Rubin Observatory \citep{Ivezic2019} would be useful for testing the jet structure.

\section{Conclusion}
\label{sec:summary}
In this paper, we have investigated a two-component jet model to explain the multiwavelength afterglows of VHE GRBs~201216C and 221009A.
For GRB~201216C, the two-component jet model given by \citet{Sato2023a} can explain the VHE flux without changing model parameters.
In GRB~221009A, our narrow jet well describes the observed peak and break in the early-time VHE light curve,
and the late optical, X-ray and HE gamma-ray light curves are consistent with the emission from our wide jet.
VHE GRBs tend to have a large isotropic-equivalent gamma-ray energy \citep[e.g.,][]{Yamazaki2020}, whereas their collimation-corrected gamma-ray energy has a typical value due to the jet with a small opening angle.
Furthermore, our modeling reveals that the CBM density and the microphysical parameters vary among VHE gamma-ray events, while the values of the collimation-collected kinetic energy and the Lorentz factor are almost similar for the narrow and wide jets (see Table~\ref{tab:GRB_model}).
Since the emission from the narrow jet decays rapidly, the late X-ray and optical afterglows require a wide jet component.
In the future, if dip and/or bumps are seen in observed light curves, 
it suggests the existence of distinct two jet components, although such features may be smoothed out in general structured jet models.
The VHE gamma-ray afterglow emission gives us a strong constraint on the modeling.

\section*{Acknowledgements}
We thank 
Katsuaki~Asano,
Tsuyoshi~Inoue,
Asuka~Kuwata,
Koutarou~Kyutoku, 
Koji~Noda and
Shuta~J.~Tanaka
for valuable comments.
We also thank the anonymous referee
for his/her/their helpful comments that improved the paper.
This research was partially supported by JSPS KAKENHI Grant 
Nos.~25KJ0010 (YS), 20H05852 (KM), 
21H04487 (YO), 24H01805 (YO), 25K00999 (YO),
23K22522 (RY), 23K25907 (RY) and 23H04899 (RY).
The work of K.M. is supported by the NSF Grant Nos. AST-2108466, AST-2108467 and AST-2308021. 

\typeout{}

\bibliographystyle{elsarticle-harv} 

\bibliography{GRB}

\end{document}